 \documentclass[journal, comsoc]{IEEEtran}

\usepackage{gensymb}
\usepackage{cite}
\usepackage{amsmath,amssymb,amsfonts}
\usepackage{graphicx}
\usepackage{xcolor}
\usepackage{kotex}  

\usepackage{caption}
\usepackage{subcaption}

\usepackage{stfloats}
\usepackage{float}
\usepackage{wrapfig}
\usepackage{xcolor}
\usepackage{colortbl} 
\usepackage{color,soul}
\usepackage{url}
\usepackage{verbatim}
\usepackage{graphicx}
\usepackage{cite}
\usepackage{multirow}
\usepackage{array,boldline, makecell,booktabs}
\usepackage{longtable} 
    \usepackage[group-separator={,},group-minimum-digits=4]{siunitx}

\begin{document}
\title{Integrated Sensing and Communications for IoT: Synergies with Key 6G Technology Enablers}

\author{Aryan Kaushik, Rohit Singh, Ming Li, Honghao Luo, Shalanika Dayarathna‬, Rajitha Senanayake, \\ Xueli An, Richard A. Stirling-Gallacher, Wonjae Shin, and Marco Di Renzo
%, and H. Vincent Poor 
\thanks{A. Kaushik is with the School of Engineering \& Informatics, University	of Sussex, UK (e-mail: aryan.kaushik@sussex.ac.uk). \\$~~~$R. Singh is with the Department of Electronics \& Communication Engineering, Dr. B. R. Ambedkar National Institute of Technology, India (e-mail: rohits@nitj.ac.in). 
\\$~~~$M. Li and H. Luo are with the School of Information and Communication Engineering, Dalian University of Technology (e-mail: \{mli,luohong-hao\}@dlut.edu.cn).
\\$~~$S. Dayarathna and R. Senanayake are with the Department of Electrical \& Electronic Engineering, University of Melbourne, Australia (e-mail: \{s.dayarathna,rajitha.senanayake\}@unimelb.edu.au). 
%\\$~~~~$J. A. Zhang is with the School of Electrical and Data Engineering, University of Technology Sydney, Australia (e-mail: andrew.Zhang@uts.edu.au).
\\$~~~$X. An and R. A. Stirling-Gallacher are with the Huawei Technologies, Germany (e-mail: \{xueli.an, richard.sg\}@huawei.com). 
%\\$~~~$H. Zhang is with the School of Electronics, Peking University, China (e-mail: hongliang.zhang@pku.edu.cn).
\\ $~~~$W. Shin is with the School of Electrical Engineering, Korea University, South Korea, and the Department of Electrical Engineering, Princeton University, USA (e-mail: wjshin@korea.ac.kr).
\\$~~~$M. Di Renzo is with Universit\'e Paris-Saclay, CNRS, CentraleSup\'elec, Laboratoire des Signaux et Syst\`emes, 3 Rue Joliot-Curie, 91192 Gif-sur-Yvette, France. (e-mail: marco.di-renzo@universite-paris-saclay.fr).  
%\\$~~~$A. L. Swindlehurst is with the Center for Pervasive Communications and Computing, University of California, Irvine, USA (e-mail: swindle@uci.edu). 
%\\$~~~$O. A. Dobre is with the Faculty of Engineering and Applied Science, Memorial University of Newfoundland, Canada (e-mail: odobre@mun.ca).
%\\$~~~$H. V. Poor is with the Department of Electrical Engineering, Princeton University, USA (e-mail: poor@princeton.edu).
%(\textit{Corresponding Author: W. Shin})
}}% <-this % stops a space

% and the end of the author  line. i.e., if you had this:

%\markboth{Submitted to IEEE Internet of Things Magazine}%
%\markboth{DRAFT, 2023}%

\maketitle

%\vspace{-15cm}
\begin{abstract}
The Internet of Things (IoT) and wireless generations have been evolving simultaneously for the past few decades. Built upon wireless communication and sensing technologies, IoT networks are usually evaluated based on metrics that measure the device's ability to sense information and effectively share it with the network, which makes Integrated Sensing and Communication (ISAC) a pivotal candidate for the sixth-generation (6G) IoT standards. This paper reveals several innovative aspects of ISAC from an IoT perspective in 6G, empowering various modern IoT use cases and key technology enablers. Moreover, we address the challenges and future potential of ISAC-enabled IoT, including synergies with Reconfigurable Intelligent Surfaces (RIS), Artificial Intelligence (AI), and key updates of ISAC-IoT in 6G standardization. Furthermore, several evolutionary concepts are introduced to open future research in 6G ISAC-IoT, including the interplay with Non-Terrestrial Networks (NTN) and Orthogonal Time-Frequency Space (OTFS) modulation.
\end{abstract}


\IEEEpeerreviewmaketitle

%\vspace{-3mm}

\section{Introduction}
\label{sec:1}

The Internet of Things (IoT) has revolutionized the way we interact with our environment and technology. IoT devices rely on advanced communication protocols and networks to share the acquired data in real time. Leveraging the potential of advanced sensing and communication abilities, significant efforts are being made to revolutionize the IoT experience further. For instance, the radio communication division of the International Telecommunication Union (ITU-R) already drafted new recommendations for International Mobile Telecommunication 2030 IMT-2030 (6G) \cite{imt}. Integrated Sensing and Communication (ISAC), one of the key 6G usage scenarios, is of particular interest since it enables one to sense and better understand the physical world. ISAC embodies the seamless fusion of two critical IoT components: sensing capabilities and communication infrastructure over a single integrated platform or cooperating as individual entities over a single network and hence complements various modern 6G IoT use cases.

\begin{figure}[t!]
    \begin{center}    %\vspace{-2mm}
        \includegraphics[width=0.9\linewidth]{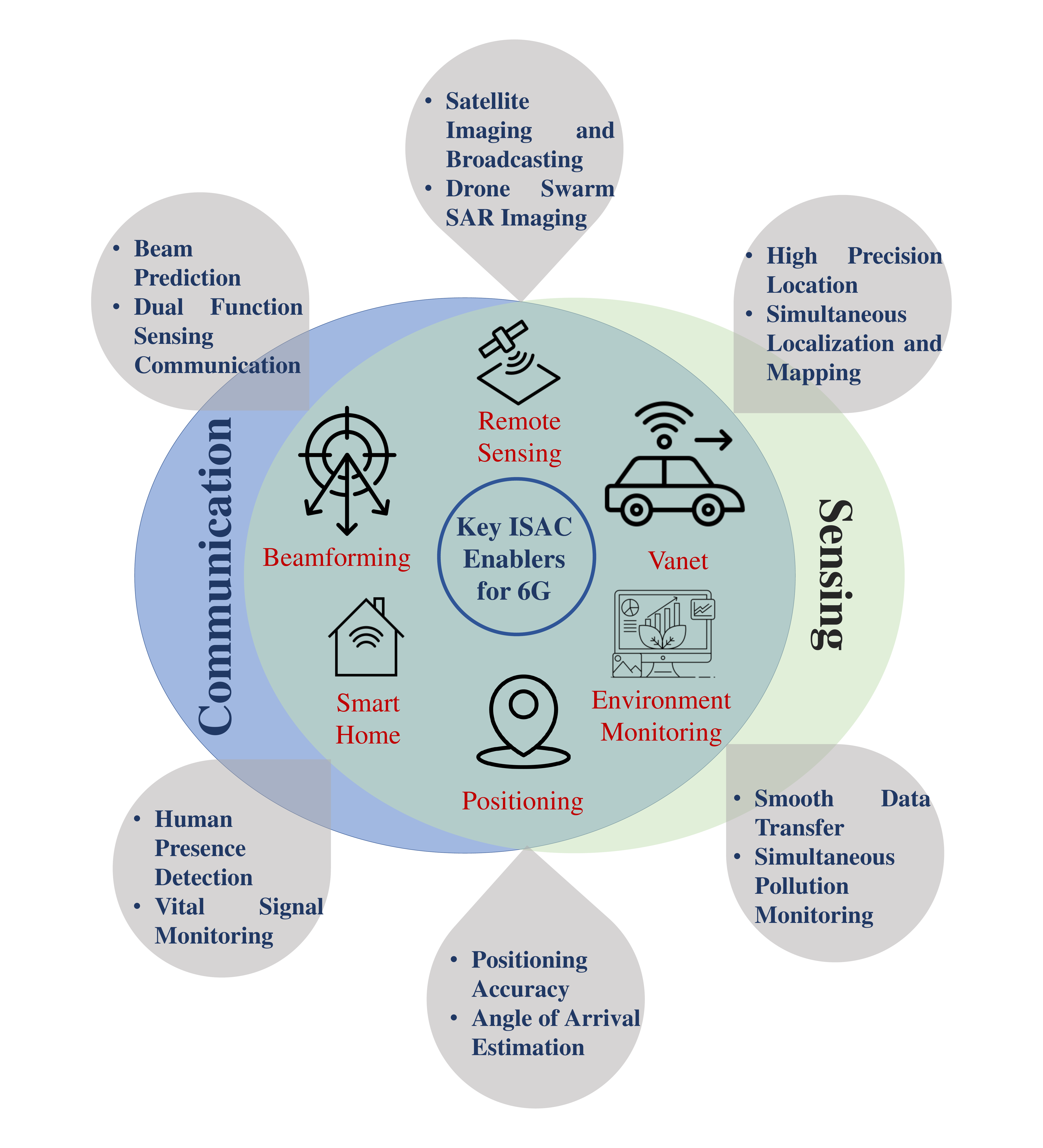}
    \end{center}
    \setlength{\belowcaptionskip}{-2pt}%\vspace{-1mm}
    \caption{ISAC enabled 6G IoT framework.}
   
    %\vspace{-3mm}
    \label{fig1a}
\end{figure}

%Indeed, 

An illustration of the components of ISAC-enabled IoT is given in Fig. \ref{fig1a}, which covers almost a complete 360$^0$ IoT framework, including enhanced positioning, seamless connectivity, smooth monitoring, and several other possibilities. In particular, ISAC offers numerous technological advantages, including \cite{isac}: \textit{a) Real-time Data Insights:} By combining sensing and communication operations, ISAC empowers IoT devices to instantaneously transmit and receive data, making it more convenient for timely decision-making,  \textit{b) Predictive Analytics:} Joint utilization of emerging processing tools and real-time data analysis enables IoT applications to predict trends, anomalies, and potential issues, \textit{c) Personalized Experiences:} IoT devices with ISAC can tailor experiences as per the individual user's requirements and location. 

\begin{figure}[t!]
    \begin{center}    %\vspace{-2mm}
        \includegraphics[width=1\linewidth]{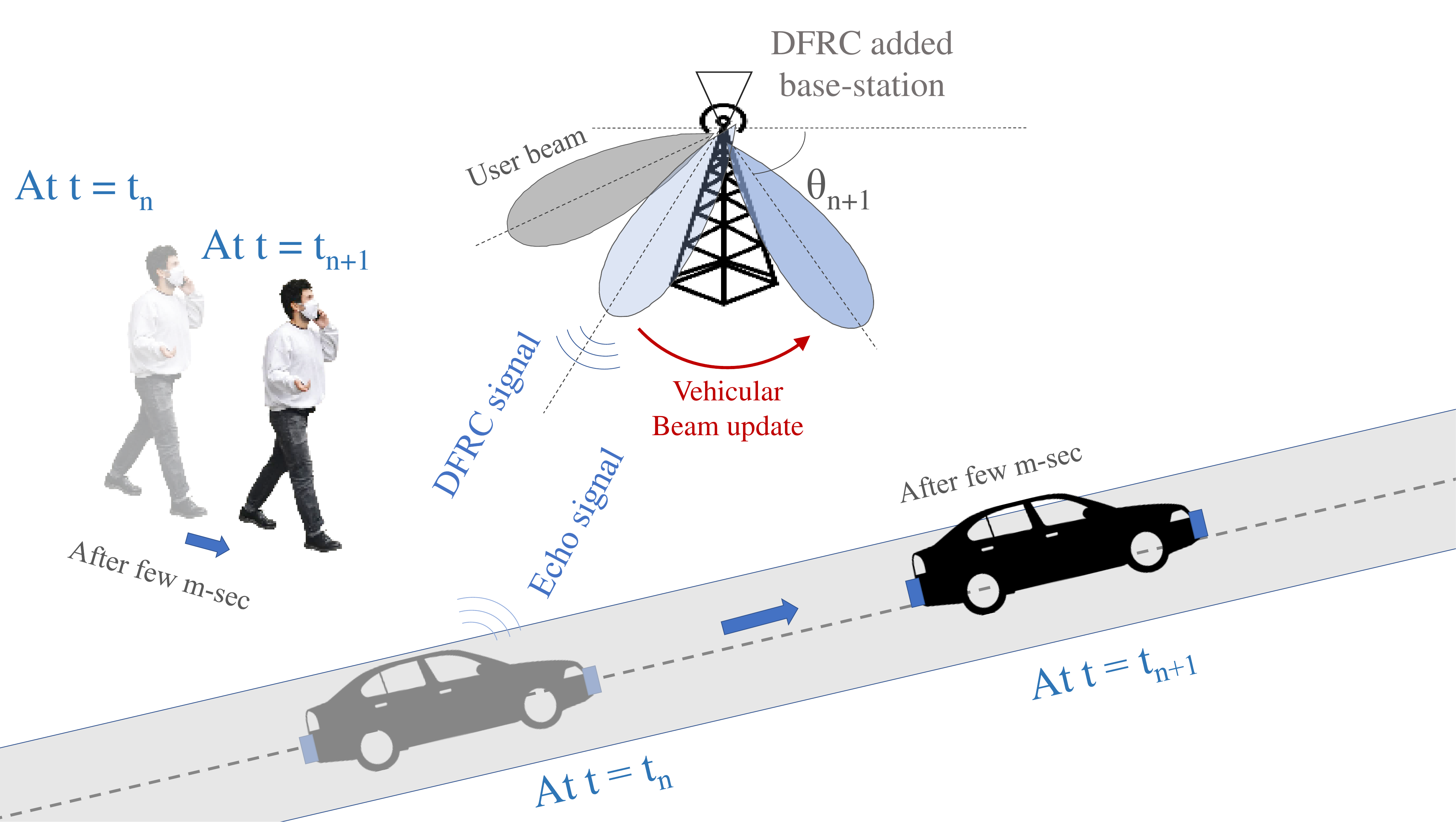}
    \end{center}
    \setlength{\belowcaptionskip}{-2pt}%\vspace{-1mm} 
    \caption{DFRC-based ISAC for predictive beamforming.}
   %\vspace{-3mm}
    \label{fig2a}
\end{figure}

The coexistence of ISAC, whether on a single platform or as individual sensing and communication units cooperating with each other, can provide most of the required IoT features envisioned in the next generation of wireless services. For instance, enabling real-time data insights makes it more convenient to support critical IoT applications such as smart cities. All such features pave the way for innovative applications across industries, transforming the way we interact with technology. Given the emergence of 6G-IoT and the novel features of ISAC, this work highlights the benefits of ISAC implementations in 6G IoT, including its key enabling techniques, use cases, challenges, and future prospects. 

\section{ISAC-enabled IoT: Benefits and Integration}
\label{sec:2}

%\subsection{Key Enabling Features of ISAC-enabled IoT}
\subsection{Transformative Benefits of ISAC}

\textit{Better Resource Management:} ISAC enables communication and sensing functionalities on a common platform, providing better spectrum utilization to support ultra-low latency and high bandwidth connections. This is indeed crucial for most of the recent IoT applications, i.e., industrial automation, autonomous vehicles, and augmented reality, etc.

\textit{Massive Connectivity:} 6G and Beyond (6GB) networks aim towards a massive number of IoT devices where ISAC can play a key role through accommodating diverse deployment needs, ranging from smart cities to healthcare, agriculture, smart transport, environmental monitoring and public safety applications.

\textit{Contextual Awareness and Cost Efficiency:} ISAC allows IoT devices to gather and process rich contextual information, especially gathered from their surroundings. An optimal utilization of these functionalities provides several benefits, including better decision-making processes, intelligent responses, etc., such as, for example, a reduction in frequent high-power communication, extending the battery life of devices.

\textit{Distributed Intelligence and Edge-Cloud Computing:} The coexistence enabled by ISAC allows IoT devices to collaborate and share device-to-device information directly with each other. This enables the networks to make decentralized decisions at the device end itself.

\textit{Security and Privacy:} ISAC can also help in enhanced security through localized threat detection and anomaly recognition. In addition, the integration of communication and sensing can provide enhanced user privacy, e.g., minimizing the transmission of sensitive information. 

\subsection{ISAC Support for Critical IoT Applications}
6G is intended to support critical IoT applications mainly characterized by the degree to which they fulfill various metrics, including reliability, resilience, sustainability, responsiveness, accuracy, etc. In the following, we summarize how the joint utilization of communication and sensing fulfills the key aspects of critical IoT applications.

\textit{ISAC As DFRC:}
Under the umbrella of ISAC, Dual Function Radar Communication (DFRC) also proposes sensing and communication functionalities on a single platform \cite{dfrc}, where the sensing function involves radar probing and target mapping. DFRC can achieve simultaneous data transmission and localization of a target source in terms of its signal's estimated Angle of Arrival (AoA). As depicted in Fig. \ref{fig2a}, DFRC is helpful in millimeter Wave (mmWave) systems, especially in high mobility scenarios, where it can provide seamless beam estimation and prediction to support critical applications such as those in vehicular networks.     

\textit{ISAC Meets Intelligent Metasurfaces:}
Another parallel research thrust focuses on the fusion of Reconfigurable Intelligent Surfaces (RIS) with ISAC to support a seamless integration \cite{iris}. RIS is known for its ability to control the transmission environment via tuning multi-path phases, which makes it a convenient 
tool in supporting ISAC applications. RIS-added benefits to ISAC are discussed in more detail in subsequent sections.

\begin{figure}[t!]
    \begin{center}    %\vspace{-2mm}
        \includegraphics[width=1\linewidth]{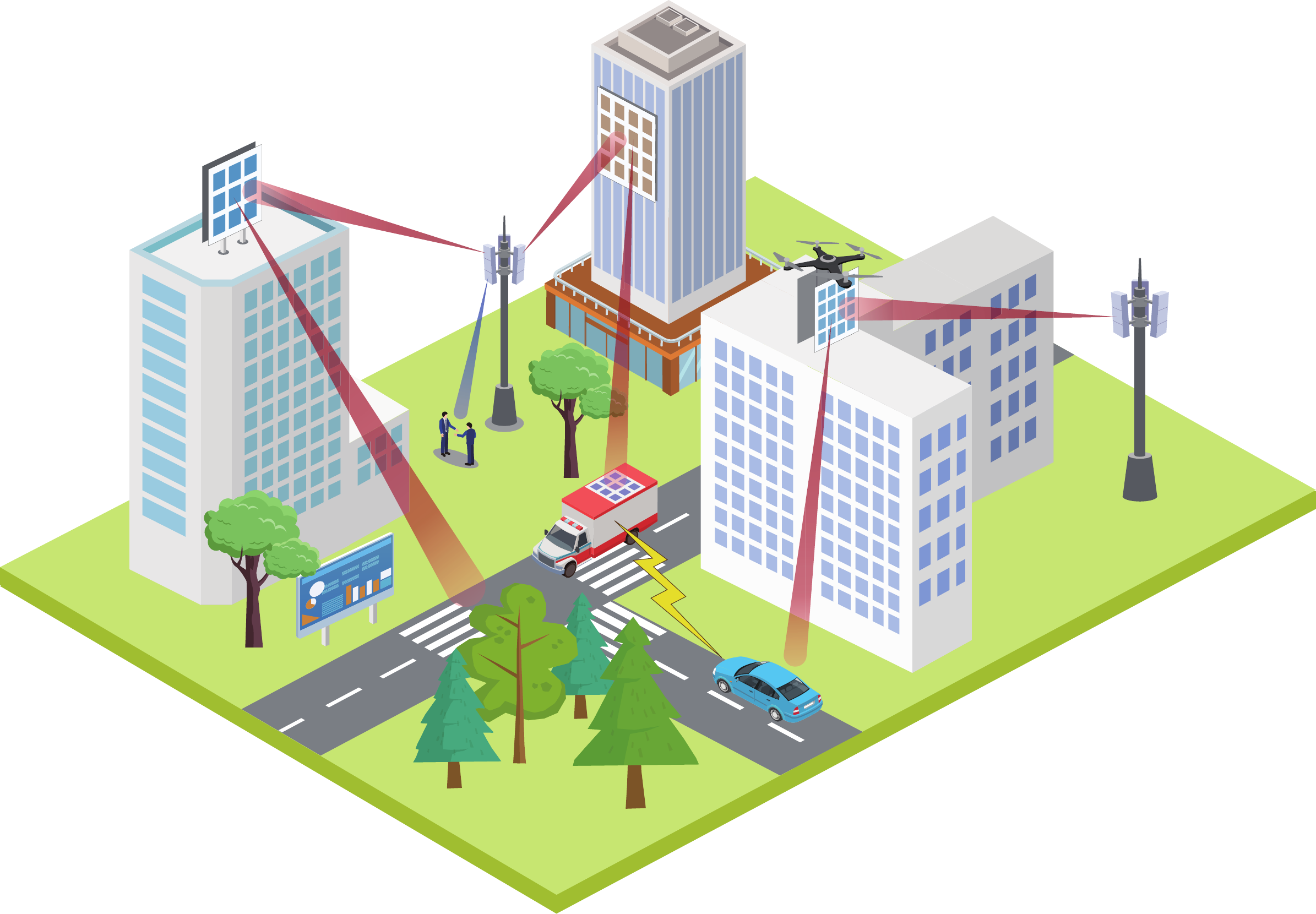}
    \end{center}
    \setlength{\belowcaptionskip}{-2pt}%\vspace{-1mm}
    \caption{RIS deployments in the ISAC-IoT network.}
    %\vspace{-3mm}
    \label{Fig:RIS_ISAC_IoT}
\end{figure}

\section{Intelligent Metasurfaces and AI for ISAC-IoT}
\label{sec:3}

\subsection{RIS-aided ISAC-IoT}
\label{sec:3.1}
Owing to the superior ability to construct an intelligent and controllable electromagnetic environment by manipulating the electromagnetic 
properties of the metasurface, RIS has been envisioned as a promising technique to revolutionize ISAC-IoT networks. As depicted in Fig. \ref{Fig:RIS_ISAC_IoT}, RIS can be deployed in ISAC-IoT networks to enhance the connectivity and coverage for various practical applications, e.g., intelligent transportation, smart cities, and beamforming \cite{ibeam}. The fusion of ISAC-IoT and RIS is more than the combination of two functionalities. ISAC-IoT grants devices not only enhanced communication performance but also active participation in the realm of environmental sensing. 

The deployment of RIS introduces a layer of intelligence to ISAC-IoT networks, endowing them with the capacity for real-time signal enhancement, energy-efficient and reliable transmission and sensing, connectivity and coverage expansion, and effective interference management. In addition, with the ability to manipulate electromagnetic waves, intelligent metasurfaces can also act as wireless sensors, capturing environmental data such as temperature, humidity, and even the presence of objects, providing IoT devices with a comprehensive understanding of their surroundings. Therefore, the deployment of RIS offers much more than incremental performance improvements; plus, it plays a vital role as an active contributor to ISAC-IoT networks.

\begin{figure}[t!]
    \begin{center}    %\vspace{-2mm}
        \includegraphics[width=0.92\linewidth]{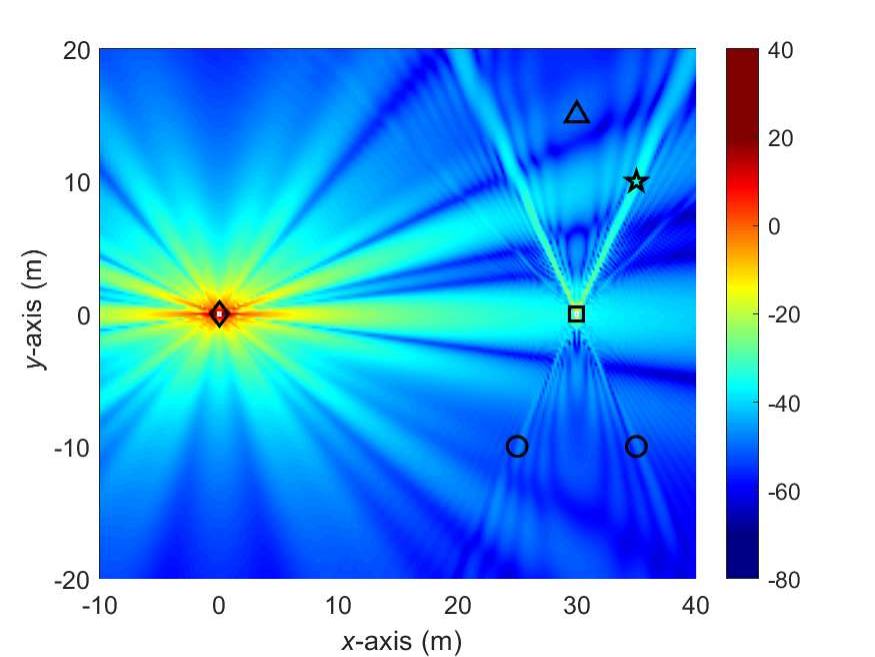}
    \end{center}
    \setlength{\belowcaptionskip}{-2pt}%\vspace{-1mm}
    \caption{Beampattern of RIS-assisted ISAC-IoT system (BS: diamond, RIS: square, target: star, clutter: triangle, user: circle).}
   % \vspace{-3mm}
    \label{Fig:beampattern}
\end{figure}

\subsection{Artificial Intelligence for RIS-aided ISAC-IoT}
\label{sec:3.3}
As discussed above, the joint utilization of RIS with ISAC can further extend various new capabilities within the realm of IoT. RISs are designed to go beyond passive functionality by incorporating sensors, communication modules, and Artificial Intelligence (AI)-driven processing to enable a wide range of advanced IoT applications. Indeed, RIS can be further fused with AI-driven algorithms, making it convenient to support modern IoT applications. In the following, we cover some of the key aspects of AI-driven RIS-aided ISAC-IoT: 

\textit{AI Assisted RIS Training for Massive IoT:}
RIS provides signal enhancement at the receiver but can require high-complexity channel training. Effective utilization of AI models, especially based on modern Machine Learning (ML) algorithms like deep neural networks, are intended to provide smooth formation of relationships between transmitted signals, reflections, and received signals. The model can be trained using the collected data from various IoT nodes to predict the channel characteristics, such as path loss, phase shifts, and delays, for different scenarios. 

\textit{ML Based Pre-coding and Optimization:} Similar to channel training, RIS transmission precoding and optimization involve extensive computational complexity due to numerous multi-paths and channel coefficients, thereby scaling drastically with the advent of massive IoT. Nevertheless, ML-based solutions can also be applied to learn complex relationships and patterns through various RIS elements, enabling more effective precoding strategies and optimization.    

\textit{AI Added Data Security:} The evolution of massive IoT is accompanied by another issue in the form of data security, especially in critical IoT use cases requiring sensitive data sharing. Research efforts are being conducted to combat such issues. For instance, AI-augmented data security pillared on RIS involves utilizing innovative AI techniques for the manipulation of electromagnetic signals for better security. 

\begin{figure}[t!]
    \begin{center}    %\vspace{-2mm}
        \includegraphics[width=0.85\linewidth]{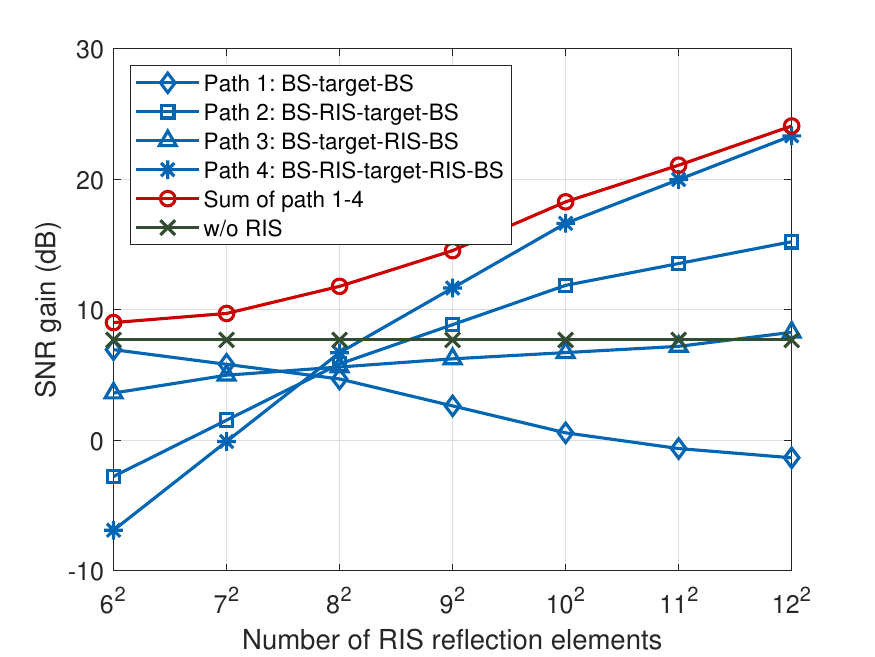}
    \end{center}
    \setlength{\belowcaptionskip}{-2pt}%\vspace{-1mm}
    \caption{Radar SNR gain versus the number of RIS reflection elements.}
 %   \vspace{-3mm}
    \label{Fig:SNR_N}
\end{figure}

\subsection{Demonstrating RIS-aided ISAC-IoT: Results and Future Prospects}
\label{sec:3.3}
In order to better reveal the potential and advantage of deploying RIS in ISAC-IoT systems, a case study is provided in this subsection. We consider an RIS-assisted ISAC-IoT system, where a colocated multi-antenna Base Station (BS) or Access Point (AP) simultaneously performs multi-user communications and target detection with the assistance of an RIS. In particular, the BS/AP simultaneously transmits data to two single-antenna users and detects one target in the presence of a clutter source. Utilizing the algorithm in \cite{Liu2022JSTSP}, we aim to maximize the radar Signal-to-Interference-plus-Noise Ratio (SINR) and satisfy the communication Quality-of-Service (QoS) and constant-modulus transmit power constraints.

We first present the resulting beampattern in Fig. \ref{Fig:beampattern}. We display the signal power at each location in different colors according to the colorbar shown on the right-hand side in Fig. \ref{Fig:beampattern}. In the considered system, the channel condition of the direct links is worse than that of reflected links. Therefore, the beams generated by the BS are mainly transmitted toward the RIS, whose passive beams further assist the BS to serve users and detect the target.

We plot the radar signal-to-noise ratio (SNR) gain of different paths versus the number of RIS elements $N$ in Fig. \ref{Fig:SNR_N}. The scheme without an RIS (“w/o RIS”) is also included for comparison. It can be easily observed that an RIS with more reflection elements offers higher radar SNR as it provides more spatial Degrees-of-Freedom (DoF). Besides, as the number of RIS elements increases, the BS will generate a stronger beam toward the RIS and a weaker beam toward the target to achieve better sensing performance because the virtual Line-of-Sight (LoS) link established by the RIS is enhanced. In particular, the role of the LoS link between the BS and the target dominates for relatively small $N$, while the virtual LoS link created by the RIS is important for large $N$.

In order to stimulate the deployment of RIS in ISAC-IoT systems and capture the opportunities they offer, we discuss several opportunities and shed light on various future directions, such as: 

\textit{Multiple RISs:} Since multiple RISs can offer additional DoFs, their geographical distribution presents opportunities for optimizing signal transmission and mitigating interference. By strategically deploying multiple RISs, we can achieve high-quality communication QoS and accurate sensing performance in hotspots owing to the improved beamforming gains. Moreover, multiple views of a target provided by multiple RISs increase the spatial diversity of radar echo signals and thus improve the target detection and parameter estimation performance.

\textit{Target/Device-Mounted RIS:} Deploying RIS on the surface of the target or IoT device can effectively assist the cooperative sensing and tracking. This approach becomes particularly valuable when dealing with a target or IoT device that needs to be known and tracked by the ISAC-IoT system. In order to achieve better beam direction or target tracking performance, RIS can be deployed on the surface of a friendly target to enhance the back-scattered signal reflected to the radar receiver, which is especially useful when the target area is small. To achieve this goal in a complicated and dynamic electromagnetic environment, key challenges include mathematical modeling of the system, optimization of RIS reflection coefficients and design of RIS control signaling, etc.

 \subsection{Holographic MIMO-Aided ISAC-IoT }
 \label{sec:3.2}
 Holographic MIMO (HMIMO) is another groundbreaking technology that represents a software-defined transceiver capable of producing three-dimensional (3D) beams using holographic principles and is being considered a promising technique for the upcoming wireless generation. Specifically, HMIMO-aided ISAC refers to a framework that makes a joint utilization of holography MIMO systems towards enhanced sensing and communication capabilities \cite{9724245}. Since the IoT ecosystem pillars connectivity and data exchange, the incorporation of HMIMO further adds a layer of dynamic adaptability. Devices equipped with HMIMO can dynamically adjust their beams to ensure seamless communication even in challenging scenarios, such as dense urban areas.

 Bringing together holographic MIMO and ISAC empowers communication capabilities to support various modern applications, including smart city, virtual/augmented reality, industrial automation, etc. Besides, beyond communication, HMIMO-assisted IoT is also intended to revolutionize sensing abilities. Via this integration, the same beams employed for communication can be repurposed for precise sensing, opening the doors for seamless IoT deployment for various modern applications, including environment monitoring, traffic flow patterns, etc. 

\section{ISAC-IoT Standardization and Use Cases}
\label{sec:4}

\subsection{ISAC-IoT: 3GPP Standardization}
In terms of the current standardization within the 3rd Generation Partnership Project (3GPP), the Service and System Aspects working group 1 (SA1) for the Fifth Generation (5G) New Radio (NR) has conducted a Study Item on use cases and requirements for future enhancement of 5G systems \cite{3GPP22} to support ISAC sensing services for various applications. This 3GPP SA1 study has identified use cases with requirements for Vehicular-to-X (V2X), Unmanned Aerial Vehicles (UAVs), 3D map reconstruction, smart city, smart home, factories, healthcare, and applications for the maritime sector \cite{3GPP22837}. Within the indoor factory setting and of particular relevance to ISAC for Industrial IoT (IIoT), the following use cases are considered: detection of Automatic Guided Vehicles (AGV) and measurement of proximity to humans, collision avoidance for Autonomous Mobile Robots (AMR) and AGVs, the combined use of sensing and localization for accurate positioning of AGVs and AMRs, gesture detection and recognition, etc.

This 3GPP Study Item paves the way for future specifications of the Service and System Aspects (SA) and the Radio Access Network (RAN) required to support sensing services in subsequent releases (e.g., Release 19 and beyond) within the constraints of 5G systems. It is, however, expected that 6G will be designed from the outset to support ISAC in a more optimized way natively. Therefore, the expectation is that 6G will support ISAC-IoT in further use cases with more demanding Key Performance Indicators (KPIs).   

\subsection{ISAC-IoT Use Cases: Potential for 6G and Beyond}
\textit{High-Accuracy Localization and Tracking:} In a 6G ISAC-IoT system, the large bandwidth beyond mmWave bands and ultra-massive MIMO technologies can provide superior resolution and excellent multi-path resolving capabilities for high-accuracy localization applications for both device-based (e.g., UE connected in a 6G network) and device-free (environmental objects) scenarios. In addition, the dense deployment of massive antennas will also enable high-precision direction estimation. With such sensing capabilities, collaborative robots in future smart factories can work safely with humans and also with each other. Some examples of collaboration between moving robots may include a drone landing on a moving carrier vehicle to get charged; a delivery robot filling a smart container (bin/tank) when detected as empty, etc. In such proximity use cases, centimetre-level localization accuracy is required to perform the tasks.

\textit{Simultaneous Imaging, Mapping, and Localization:} 6G ISAC-IoT based sensing capabilities in simultaneous imaging, mapping, and localization enable mutual performance improvements for these functions, which opens up a realm of possibilities in 3D indoor/outdoor non-line-of-sight imaging and mapping. For instance, the sensors on a single mobile vehicle usually have a restricted view and limited coverage due to the weather, obstacles, and the sensors’ power control. That said, nearby moving vehicles and stationary base stations can jointly provide a greater ﬁeld of view, longer sensing distance, and higher resolution with the help of an ISAC system. 

Thus, the vehicles can use the reconstructed map processed by the base stations to determine their next move to achieve higher levels of autonomy. Furthermore, the sensing resolution and accuracy signiﬁcantly improve due to the fusion of imaging results that are shared globally through the network with cloud-based services. Densely distributed base stations in an urban area, together with ISAC, make environmental reconstruction and 3D localization possible, which in turn can be used to form a virtual urban city. Similar use cases could be applied to indoor factories with many autonomously moving AGVs or robots.

\textit{Augmented Human Sense:} With the use of much higher frequency bands, an ISAC-IoT system can be implemented in portable devices to augment human senses and enable people to \emph{see} beyond the limits of human eyes. Such capabilities can be incorporated into portable terminals, e.g., sensing devices such as 6G-enabled mobile phones, wearables, or medical equipment implanted beneath the human skin. This will open the door for numerous applications such as remote surgery, detection of product defects, sink leakage detection, etc., where very high range and cross-range resolutions are required.

 \textit{Smart Healthcare:} Device-free gesture and activity recognition based on the joint capability of sensing and ML is another promising aspect of 6G ISAC-IoT applications to promote contactless user interfaces and camera-free supervision where privacy can be protected. With high classification accuracy, many functionalities, such as gesture recognition, emotion recognition, heartbeat detection, fall detection, respiration detection, sneeze sensing, intrusion detection, etc., can be implemented in a smart hospital in the foreseeable future. As a novel usage scenario, a medical rehabilitation system in a smart hospital enables automatic supervision of patients during their physiotherapy exercises. There will be automatic prompt alerts for incorrect movements or gestures, thus signiﬁcantly improving the patients’ rehabilitation. 
 %In the following, we present the interplay of NTNs with ISAC-IoT.

% \subsection{\textcolor{red}{ISAC-IoT 6G and Beyond Use Cases [1/2 page] (Xueli+Richard+Aryan+Rohit)}} [almost same use-cases are being added in other sections, may be added later if required]
% ISAC allows IoT devices not only to collect data but also to communicate the same in real-time, enabling more sophisticated applications and services. Especially for 6G and beyond, ISAC coexistence intends to benefit in various ways including:

%\textit{Smart Healthcare and Cities:} ISAC enabled IoT devices will be able to capture data from several sensors including patients' health conditions in real-time, traffic flow, waste management, and public safety.  
%Moreover, 6G added capabilities enable on chip processing for rapid reaction towards various uses cases including remote healthcare, optimizes urban planning, etc.

\subsection{ISAC IoT Over Non-Terrestrial Networks}
As the number of IoT sensors grows exponentially, achieving massive connectivity based on traditional terrestrial infrastructure becomes challenging. Moreover, IoT sensors in remote areas such as mountains, forests, oceans, deserts, and rural regions are difficult to serve through existing Terrestrial Networks (TN). To overcome these issues, low Earth orbit (LEO) satellite mega constellation networks with low latency and high throughput can be promising solutions. Moreover, the integration of UAVs and high altitude platform stations (HAPS), which are closer to the ground-based IoT sensors, enable the creation of a more flexible Non-Terrestrial Network (NTN) \cite{vaezi2022cellular}. 
%This results in improved connectivity and flexibility of NTN-IoT systems. 

\begin{figure}[t!]
    \begin{center}    %\vspace{-2mm}
        \includegraphics[width=0.85\linewidth]{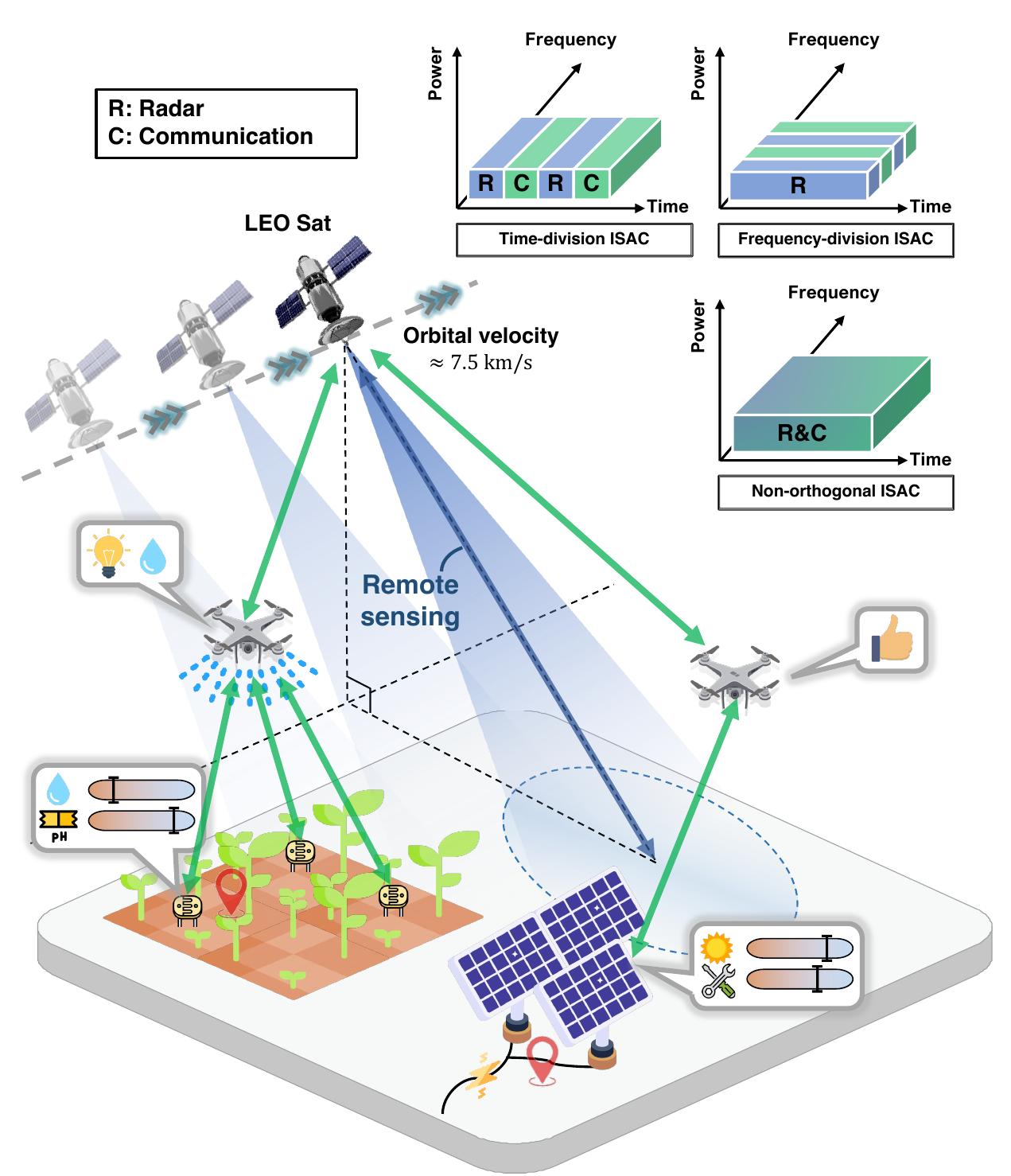}
    \end{center}
    \setlength{\belowcaptionskip}{-2pt}%\vspace{-1mm}
    \caption{NTN and its interplay with ISAC-IoT.}
    %\vspace{-3mm}
    \label{NTN_ISAC_IoT}
\end{figure}

\textit{ISAC-IoT Meets NTN:} The interplay of NTN with ISAC-IoT provides an opportunity that enables NTN platforms to serve both communication and sensing functions. As shown in Fig. \ref{NTN_ISAC_IoT}, this integration enables NTN platforms not only to seamlessly communicate with a large number of IoT sensors in remote areas but also to simultaneously perform remote sensing functions such as synthetic aperture radar (SAR) \cite{wang2019first}. In order to achieve both massive connectivity and satisfactory sensing performance by leveraging the wide and flexible coverage of NTN, efficient resource allocation and multiple access techniques are vital. As depicted in Fig. \ref{NTN_ISAC_IoT}, satisfying these demands through orthogonal resource allocation methods like time-division and frequency-division ISAC could be challenging due to the inefficient use of the wireless resource.  In this regard, overlapped resource allocation methods based on non-orthogonal multiple access (NOMA) and Rate-Splitting Multiple Access (RSMA) are promising candidates. In this case, joint waveform design and interference control techniques become inevitable to simultaneously meet both communication and sensing performance requirements at a satisfactory level.

\textit{UAV-Mounted RIS for ISAC-IoT:} Combining the high mobility and flexibility of UAVs with the low cost, light weight and low power consumption of RIS, creates an excellent means for providing reliable coverage in ISAC-IoT systems. Lightweight RIS can be mounted on the UAV to create virtual LoS links to cover shadowed areas. Nevertheless, it is crucial to acknowledge that signals reflected from the UAV may interfere with target detection, and Doppler frequency components generated by the moving UAV may introduce ambiguities in velocity measurements. Therefore, the joint design of RIS reflection coefficients associated with transmit beamforming, received signal processing, and the UAV trajectory becomes an important tool for achieving satisfactory communication and sensing performance.

%\section{\textcolor{red}{ISAC-OTFS Waveform Optimization [Overall, around 1 + 1/2 pages]}}
%\label{sec:4}
\subsection{ISAC-IoT: Facilitating OTFS Modulation}
\label{sec:4.1}
Communication in high-frequency bands provides the most efficient method to improve the capacity of IoT networks. As a result, mmWave and terahertz bands are becoming more popular in future IoT devices. This makes future IoT devices more susceptible to the impact of non-ideal power amplifiers and phase noise in high-frequency bands. Similarly, high-mobility IoT applications suffer from high Doppler due to complex time- and 
frequency-variant channels. Therefore, existing solutions fail in the context of future IoT devices that utilize high-frequency bands and operate in high-mobility networks. 

Orthogonal Time Frequency Space (OTFS) modulation is a novel approach that provides an attractive solution for this problem due to its resilience to time and frequency-varying channels, low Peak-to-Average-Power Ratio (PAPR), robustness to phase noise, and higher spectral efficiency \cite{3132606}. OTFS modulation-based solutions have been proposed for several IoT application areas, such as future vehicular networks, underwater acoustic communication, and NTN. For example, integrated grant-free non-orthogonal multiple access with OTFS is proposed to mitigate the severe Doppler effect and round trip delay experienced by LEO satellite-based non-terrestrial IoT networks.% \cite{3214862}. 
Given the advantages of using ISAC in IoT networks, it is reasonable to consider OTFS as a potential waveform for IoT-ISAC networks. Existing research on OTFS-ISAC enabled IoT focuses on the following topics:

\textit{High-Mobility Vehicular IoT Networks:} OTFS-ISAC-enabled vehicular networks allow the Road Side Unit (RSU) to obtain the necessary information to generate a dynamic vehicle network. Considering the precision of delay and Doppler shift estimation in OTFS, OTFS-ISAC signals are utilized in \cite{9968-0_64} to predict the motion state of moving vehicles combined with wide beams to estimate the position. This approach addresses the issue of vehicles with high speed not being covered by narrow beams in high-mobility vehicular IoT networks.

\textit{RIS-aided OTFS for IoT:} RIS-aided OTFS is capable of facilitating future IoT networks due to the combined benefits of flexible channel configurations and robustness in high-mobility communication. Given the real-time communication and sensing needed in future IoT networks, a new transmission scheme is proposed in \cite{3270335} that uses delay and Doppler shifts in cascaded channels for sensing the user, followed by configuration of RIS passive beamforming based on the sensed parameters. 

\textit{Low-Complexity Sensing for IoT:} Despite the popularity of OTFS-based ISAC solutions, existing sensing solutions involve complex algorithms and high computational complexity, which makes them infeasible for IoT devices. Recently, an echo pre-processing approach \cite{3139683} and new waveform designs that couple OTFS with Frequency Modulated Continuous Waveforms (FMCW) %\cite{9977883} 
have been proposed for OTFS-ISAC-aided IoT. However, research areas, such as sensing in the presence of timing and frequency offsets due to distributed, non-coherent, and asynchronous transceivers, are yet to be explored.

% \textit{Limitations of OTFS ISAC Implementation}
% The mentioned advantages offered by OTFS modulation and ISAC for IoT indicate a paradigm shift in the existing IoT solutions. Nevertheless, the limited power, computing resources, and storage available in IoT devices result in several major research problems that need to be addressed before successful implementation of OTFS-ISAC enabled IoT. These can be identified as,
% \begin{itemize}
%     \item \textbf{Low-complexity signal detector:} OTFS requires high complex maximum likelihood (ML) decoding to achieve full time-frequency (TF) domain diversity. Accordingly, it is necessary to further analyze the suitability of these low-complexity receiver designs.
%     \item \textbf{Efficient waveform design:} A rectangular pulse, i.e., existing solutions often utilize pulse-based OTFS waveform, might result in interference with adjacent channels, thus increasing the out-of-band (OoB) power emission. As such, it is important to explore the efficient waveform designs that address the OoB power emission
%     \item \textbf{Integration with other IoT solutions:} Given that several IoT solutions exist, it is important to investigate the integration of OTFS-enabled ISAC with those existing IoT solutions.
% \end{itemize}

%\subsection{\textcolor{red}{OTFS-ISAC Results (Shalanika+Rajitha)}}
% \label{sec:4.2}

\section{ISAC-IoT Challenges and Future Capabilities}
\label{sec:4}

ISAC in IoT systems provides numerous potential benefits, although innovative approaches are required to enable advanced ISAC coexistence. Nevertheless, ISAC opens up several doors for future applications, research, and integration. %Specifically, the implementation challenges and future directions for ISAC-added IoT are added below:  
\subsection{Challenges and Solutions}
\textit{Network Reliability and Security:} Wireless transmissions generally encounter impairments including attenuation, interference, noise, etc., and ISAC integration necessitates a robust network implementation. Some useful tools may include error correction codes or adaptive modulation techniques to maintain reliable connections. Another integration issue arises in the form of data security, e.g., ensuring data confidentiality, device authentication, etc. 

\textit{Compliance and Regulatory Considerations:} Another challenge arises in the form of non-uniform industrial regulations, e.g., data protection, electromagnetic compatibility, safety standards, etc. For seamless integration, ensuring the regulatory-approved behavior of IoT devices is necessary without compromising functionality or usability.

\textit{Cost Effectiveness and Power Management:} An exponential increase in IoT devices will proportionally affect power consumption, especially for ISAC functions. This requires efficient power management strategies, e.g., duty cycling, low-power components, sleep modes, etc., as well as innovative implementation and research efforts, optimizing hardware components and tracing, minimizing power consumption, embedding suitable communication modules, etc.

\textit{Data Volume Handling and Processing:} ISAC integration and evolving IoT techniques are expected to generate substantial amounts of data from a wide variety of sensors. This, in turn, requires efficient data handling and processing. Some helpful tools in this regard include data compression, filtering, aggregation, etc. In addition, research efforts are currently focused on advanced data processing techniques such as edge analytics, data fusion, etc.

%\textit{Cost-Effective Designing:} Bringing together sophisticated sensing and communication capabilities with cost-effectiveness is another technical challenge. This involves innovative implementation and research efforts, optimizing hardware components and tracing, minimizing power consumption, embedding suitable communication modules, etc.

\subsection{Future Capabilities}

\textit{Edge Computing Integration:} Forthcoming IoT systems will leverage edge computing, enabling data processing and analysis to occur closer to the data source. This involves deploying edge servers or gateways that employ powerful processors, GPUs, and FPGAs to execute complex algorithms, reducing latency and conserving network bandwidth. This integration requires the development of edge applications, containerization techniques, and orchestration frameworks to manage resources effectively.

\textit{Synergy with AI:} As discussed above in Section III.B, ISAC combined with the potential of AI will lead to significant advancements, creating devices that can make decisions in real-time based on the data they collect. From an industrial perspective, AI algorithms can analyze  sensor data to predict when maintenance is required. Similarly, AI-powered analysis of data received from various IoT sensors can enable doctors to detect early signs of diseases and even predict health issues before they become critical. The utilization of AI will help in numerous IoT applications including smart cities, supply chains, agriculture, security, smart transport, augmented logistics, environmental monitoring, public safety, etc. 

\textit{Multi-Modal Sensing Fusion:} Multi-modal sensor fusion aims to integrate data from different types of sensors/sources to create a more comprehensive and accurate understanding of the environment. Such integration will allow IoT systems to leverage various sensors to make more accurate decisions.  For instance, in a smart city application, capturing and processing data from different traffic cameras and sound sensors will provide a more accurate understanding of traffic conditions.

\textit{Swarm Intelligence and Collaborative Utilization:} Apart from the above technologies, ISAC-assisted IoT can also benefit from swarm intelligence and collective utilization of massive IoT units and sensor data. Inspired by nature, their collective utilization can significantly benefit IoT networks, e.g., achieving better scalability, adaptability, etc. For instance, IoT devices can communicate and adapt themselves via collective information, enabling them to make group decisions towards optimizing an overall objective.

%\textit{Blockchain for Security and Trust}

\section{Conclusion}
IoT has evolved the way we interact with technology, and the coexistence enabled by ISAC is intended to revolutionize the IoT experience in 6G. This paper discusses several innovative aspects of ISAC-enabled IoT in 6G. Specifically, this work has emphasized the following contributions: \textit{a)} comprehensively explored ISAC-enabled IoT, highlighting its benefits, applications, challenges and future prospects in 6G-IoT, \textit{b)} describing the progress of ISAC-IoT in 3GPP standardization for 6G standards, and vital use cases of ISAC-IoT in real world plus synergies with the key technology enablers such as RIS, AI, NTN and OTFS modulation, \textit{c)} highlighting the future potentials such as edge computing integration and multi-modal sensing fusion for ISAC-enabled 6G IoT.

% \section{Acknowledgement}

\bibliographystyle{IEEEtran}

\bibliography{IEEEabrv,BibRef}

% \vspace{-5mm}
% \begin{IEEEbiographynophoto}{}
% \vspace{10mm}
% \small

% \end{IEEEbiographynophoto}\vspace{-5mm}
\end{document}